\documentclass[preprint,twocolumn]{aastex6}
\usepackage{amsmath}
\usepackage{natbib}
\usepackage{threeparttable}

\def\msol{M$_{\odot}$}

\def\eg{{\it e.g.,}}
\def\ie{{\rm i.e.,}}

\newcommand{\be}{\begin{equation}}
\newcommand{\ee}{\end{equation}}

\begin{document}
\title{Shadows of our Former Companions: How the Single-Degenerate Binary Type Ia Supernova Scenario Affects Remnants}

\author{William J. Gray, Cody Raskin, \& J. Michael Owen}

\affil{Lawrence Livermore National Laboratory, P.O. Box 808, L-038, Livermore, CA 94550}

\begin{abstract}
Here we present three-dimensional high resolution simulations of Type Ia supernova in the presence of a non-degenerate companion. We find that the presence of a nearby companion leaves a long-lived hole in the supernova ejecta. In particular, we aim to study the long term evolution of this hole as the supernova ejecta interacts with the surrounding interstellar medium. Using estimates for the x-ray emission, we find that the hole generated by the companion remains for many centuries after the interaction between the ejecta and the interstellar medium. We also show that the hole is discernible over a wide range of viewing angles and companion masses.  
\end{abstract}

\section{Introduction}
%\listoftodos

Type Ia supernovae (SNe Ia) are very important tools in cosmology considering their standardizable light curves \citep{Pskovskii1977,Phillips1993,Hamuy1996,Phillips1999} which make them excellent standard candles \citep{Colgate1979,Branch1992}. Type Ia supernovae are thought to be the thermonuclear explosions of a carbon-oxygen white dwarf and are characterized by the lack of hydrogen in their spectra and the formation of large amounts of radioactive {}$^{56}$Ni. However, the precise mechanism for producing the explosion remains uncertain. 

SNe Ia progenitor scenarios generally fall into two categories. In the double degenerate scenario, two white dwarfs combine and detonate. This can be the result of inspiral in a close binary \citep{Pakmor2010,Pakmor2011,Pakmor2012,Dan2012,Guillochon2010,Sato2016}, or the result of a direct collision \citep{Rosswog2009,Raskin2009,Raskin2010,Loren2010}. Even though this latter arrangement requires a dense stellar environments, \eg\ globular clusters or galactic nuclei, these systems can produce a range of luminosities. 

The second category of SNe Ia progenitors is the single degenerate model. Here, the degenerate white dwarf shares a binary system with a non-degenerate companion. The white dwarf accretes gas from the companion through Roche-lobe overflow \citep{Whelan1973,Nomoto1982,Hillebrandt2000,Hillebrandt2013}.  To date, several non-degenerate candidates have been studied, from the canonical hydrogen-burning companions\citep[\eg][]{Hachisu2007} to helium-burning companions \citep{Iben1987,Ruiter2009,Wang2009,Ruiter2011}. The primary challenge for the canonical hydrogen-burning companion is to achieve an accretion rate of $\sim$10$^{-7}$$\rm M_{\odot}/yr$ which allows for a steady increase in the mass of white dwarf while avoiding mass loss from classical novae \citep{Nomoto1991}. At this rate, the white dwarf undergoes thermonuclear runaway as it grows toward the Chandrasekhar mass limit, M$_{ch}$=1.44\msol.

In the single degenerate case, much work has been done trying to identify the progenitors that lead to observable SNe Ia. In this scenario, the collision of the expanding supernova ejecta with the companion star is unavoidable. This led  \cite{Kasen2010} to calculate theoretical supernova light curves for a supernova interacting with a 1-2 \msol\ red giant star.  They found that at early times ($t<8$ days), the luminosity is dominated by the collision.  However, only viewing angles that look directly down on the companion will have prominent collision signatures; limiting detection to $\sim10\%$ of the entire population. \cite{Kutsuna2015} built upon this work and found that the expected UV signal is also dependent on the separation between the white dwarf and the companion; finding that for separations $<2.0\times$10$^{13}$cm the UV flux cannot be detected. \cite{Meng2016} found that many of the binary systems that lead to SNe Ia have separations much less than this cutoff, making the measurement of the UV emission much more difficult. 

Direct observation of the collision between and the supernova and a companion are notoriously difficult. \cite{Cao2015} found a strong but short lived ultraviolet emission from a young SNe Ia within four days of the explosion, which they suggest is the result of the collision between the supernova material and the companion star. However, this requires a separation distance of $\sim4\times$10$^{14}$cm. Given these constraints on the separation distance and viewing angle, the probability of observing this UV signal is low. 

One important effect (and possible observable signature) produced by the interaction between a companion and the supernova ejecta is the formation of a hole within the ejecta \cite[\eg][]{Fryxell1981,Marietta2000,GarciaSenz2012}. Naturally, ejecta material that interacts with the companion is slowed relative to the rest of the ejecta. This creates a ``mass shadow'' in the ejecta. The presence of this hole is in direct opposition to observations which find that supernova remnants are remarkably spherical \citep{Badenes2010}. Therefore, the ultimate evolution of the hole has important implications for the likelihood of progenitor scenarios for SNe Ia. 

%As shown below, this hole is long lived and should be evident in x-ray observations of SNe Ia.

\cite{GarciaSenz2012} simulated this interaction with a supernova arising from a 1.38\msol\ white dwarf and a 1\msol\ main sequence companion as well as its interaction with the surrounding interstellar medium using a cylindrical coordinates, axisymmetric smooth particle hydrodynamic (SPH) code. They found that the hole generated by the companion is slowly filled in by ejecta due to hydrodynamical instabilities at the edge of the hole. They also estimated the x-ray emission from the ejecta and showed that signatures of a hole in the ejecta should remain visible for an extended period. The strength of this signal depends greatly on the the viewing angle. Here we aim to build on this work by performing a suite of 3D simulations that vary the companion mass. 

In this paper, we carry out a comprehensive survey of hydrodynamics simulations designed to accurately measure the impact of close binary companion stars on SNe Ia remnants. Each simulation proceeds in two stages. First, we follow the interaction between the supernova ejecta and the companion, accounting for the formation of the ejecta hole and the matter stripped from the companion. Once the ejecta has reached homologous expansion and the mass stripped from the companion has plateaued, we stop that simulation and expand the ejecta to a point where the interaction with the interstellar medium (ISM) becomes important. We then follow the ejecta as it interacts with the interstellar medium. Estimates of the x-ray emission are used to probe the evolution of the hole during this interaction, finding that the hole persists for many centuries after the interaction with the ISM.

The structure of the paper is as follows. In \S 2, we describe our numerical setup for each of the binary systems we model, with companion masses ranging from 2--5\msol. In \S 3, we give the results of our simulations and we carry out a number of analyses with the goal of constraining possible observables that can be replicated for real remnants, and in \S 4, we present results from our resolution study and, finally, in \S 5 we discuss our conclusions.
%\todo{redo structure}

\section{Numerical Setup}

Since we aim to quantify the effects of binary evolution as well as interaction with the ISM on the asymmetries of the supernova remnant, it is imperative that the initial  conditions be accurate and consistent with those found in nature. Most critical among these conditions is the size and separation distance of the non-degenerate companion.  There are several possible candidates for the non-degenerate companion including main sequence stars \cite[\eg][]{vandenHeuvel1992,Langer2000,Han2004}, He stars \citep{Tutukov1996}, and red giants \citep{Hachisu1999,Patat2011}. Here, an evolved sub-giant is used as non-degenerate companion which donates material onto the degenerate primary. As described, the system has two constraints: the masses of the degenerate primary and the non-degenerate secondary.
%\todo{fix above, PROJECT: 40362 TASK: WCI POST DOC 25 }

In our case, we use a white dwarf mass of 1.0 \msol\ for each of the models presented here. This white dwarf mass can be converted to nuclear statistical equilibrium (NSE) with the right energetics and nuclear yields. This also avoids the complication of handling a deflagration to detonation transition. As shown in \citep{Badenes2006}, the structure of the SNR depends on the density profile of the ejecta, which in turn depends on the explosion model used. Therefore, for numerical simplicity the above white dwarf model is used. Finally, recent simulations by \citep{Moll2013} have shown that SNe Type Ia like explosions can be produced with 1 \msol\ white dwarfs.

The composition of the white dwarf is assumed to be 50\% Carbon and 50\% Oxygen, and we employ the Helmholtz equation of state with coulomb corrections  \citep{Timmes1999,Timmes2000} to compute the radius and zero temperature density profile. All other parameters, then, are a function of the zero age main-sequence (ZAMS) mass of the non-degenerate companion.

\begin{figure}
\centering
\setlength{\tabcolsep}{0mm}
\includegraphics[width=0.40\textwidth]{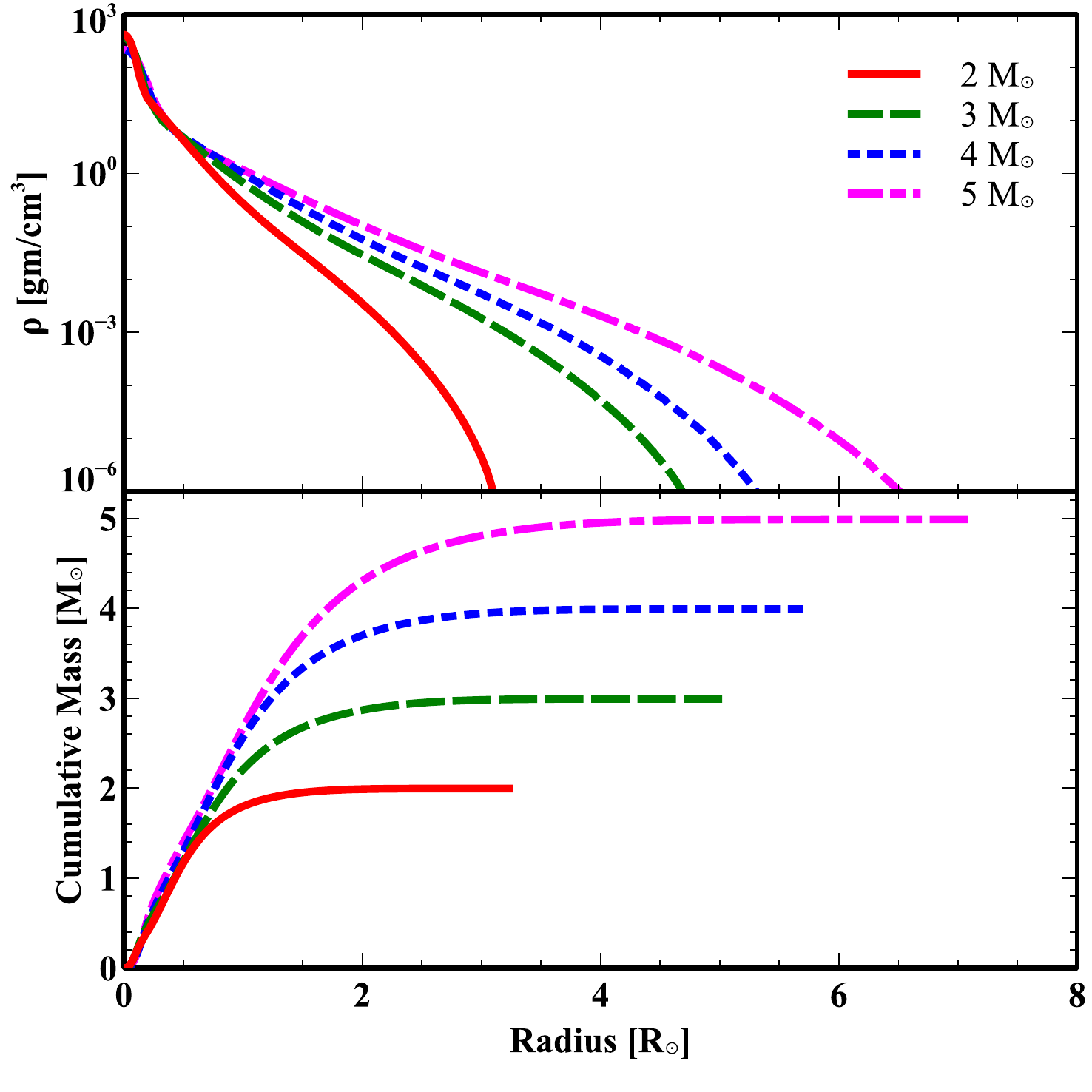} \\
\caption{ Initial conditions for the companion stars. {\it Top Panel:} Logarithm of the  mass density and {\it Bottom Panel:} the corresponding mass as a function of radius. The legend gives the corresponding line style and color for each model. }
\label{figure:ic}
\end{figure}

We calculate the radius and composition of an evolved, solar metallicity companion using the 1D stellar evolution code \textsc{mesa, rev 7624} \citep{Paxton2011}, where we assume the maximum radius of the red giant branch (RGB) phase to be the most relevant. 
Since we are aiming to simulate a ``clean'' circumstellar medium, devoid of any outbursts arising from intermittent novae, we can safely ignore the effects of prior binary evolution on the radius and density profile of the RGB and of the white dwarf.
Figure~\ref{figure:ic} shows the initial density (top panel) and mass (bottom panel) profiles as a function of radius for our companion RGB stars. 

The separation distance between the two stars is found by placing the surface of the companion at the first Lagrange point,
\be
\frac{R}{R_{*}}=\left(1.0-\left(\frac{1}{3q}\right)^{1/3}\right)^{-1},
\ee
where $R$ is the separation distance between the white dwarf and the companion, $R_{*}$ is the radius of the evolved companion, and $q=M_{\rm *}/M_{\rm wd}$ is the ratio of the companion ($M_*$) and the white dwarf ($M_{\rm wd}$) masses. 
The non-degenerate companion masses, radii, and separation distances are given in Table \ref{table:ic}. 
We also note that for this choice of separation distance, the solid angle subtended by the companion ranges between 5\% and 8\% of $4\pi$ for the 2m and 5m simulations (2\msol\ and 5\msol\ companions) respectively, larger than those found in the models of \cite{GarciaSenz2012} and \cite{Marietta2000}. 

\begin{table}[]
\centering
\caption{Summary of models.}
\label{table:ic}
\begin{tabular}{c|cccc|c|cc}
\hline
\# & $M_*$ & $M_p$ & $R_*$ & $q$ & $a$ & M$_{s}$ & M$_{s,t}$\\
\hline
\hline
5m & 5.00 & 4.4 & 7.05 & 5.00 & 11.8 & 0.11 & 0.11\\
4m & 4.00 & 3.5 & 5.66 & 4.00 & 10.0 & 0.16 & 0.16\\
3m & 3.00 & 4.4 & 5.00 & 3.00 & 9.6  & 0.18 & 0.19\\
2m & 2.00 & 4.9 & 3.23 & 2.00 & 7.2  & 0.21 & 0.22\\ 
\hline
\end{tabular}
\begin{tablenotes}{
\item \textbf{Notes.} $M_*$ and $R_*$ give the companion mass and radius in solar units. The $q$-parameter gives the mass ratio and $a$ is the initial separation in solar radii. $M_p$ gives the average particle mass of the companion in units of 6.0$\times$10$^{27}$ gm. The white dwarf has an average particle mass of 5 in the above units. M$_s$ gives the total mass stripped from the companion by the supernova in solar units. $M_{s,t}$ gives the theoretical unbound mass in solar units from \cite{Wheeler1975}. }
\end{tablenotes}
\end{table}

\subsection{SPH Setup}

All of our hydrodynamical simulations were performed in 3D cartesian using \textsc{spheral} \citep{Owen2014}, a state of the art Smooth Particle Hydrodynamics (SPH) code. 
In particular, we make use of its conservative reproducing kernel formulation \citep[CRKSPH,][]{Frontiere2016}. 
This reformulation of SPH was developed in response to some of the weaknesses in the standard SPH approach as applied to applications involving mixing.

Both the white dwarf and the companion are composed of nearly equal mass particles to within a few percent, arranged using a hybrid recursive primitive refinement and parameterized spiraling scheme \citep[RPRPS,][]{Raskin2016}, which ensures nearly identical masses for particles of varying sizes (smoothing lengths), while also maintaining a high level of spherical conformity.
This also prevents numerical artifacts from developing when two particles with a large mass difference interact \citep[\eg][]{Ritchie2001,Rosswog2009B}. 

The white dwarf is modeled with $\approx65,000$ particles while the companion is modeled with between $\approx137,000$ particles for the 2m simulation and $\approx380,000$ particles for the 5m simulation. 
Particles masses for each simulation are given in Table.~\ref{table:ic}.
These particle counts strike a balance between numerical resolution and computation efficiency. 

We use the Helmholtz free-energy equation of state \citep{Timmes1999,Timmes2000} for both the companion and the white dwarf due to its applicability to a range of gaseous states from ideal gas to degenerate electron pressure support, and including photon pressure support. 
Self gravity of the gas is also included to ensure a proper hydrodynamical response for the companion star. 

Each simulation is broken into two parts: the supernova phase and the interstellar medium (ISM) interaction phase. 
The supernova is initialized by adding a total energy of $\approx10^{51}$ ergs to the interior 25\% of the white dwarf by radius (approximating the energy input from converting the inner part of the white dwarf to NSE), which is $\sim3\times$ the total binding energy of the white dwarf. 
We use two criteria to determine when the supernova phase is completed; first we ensure that the supernova ejecta has reached homologous expansion, and second, that the amount of material stripped from the companion has converged to some maximum. 
To guarantee that the ejecta material has reached a state of homologous expansion, we compute a linear least squares fit to the particle velocities of the ejecta material. 
We define the homologous expansion phase to be the time when $R^2$, the linear regression fitting parameter, is greater than 0.95.

The second stopping criterion ensures that we have accounted for all the mass that is stripped from the companion. A particle is considered stripped when its kinetic energy is greater than its local gravitational potential, \ie, that the particle has a velocity greater than the local escape velocity. A similar criterion is used in \citep{Marietta2000}. We consider this quantity to be converged once the stripped mass varies by less than $\sim$ 0.1\% for several consecutive time steps.  

Once the ejecta has overtaken the companion and reached homologous expansion, and our two stopping criterion have been met, we cease the simulation and save the final state of the ejecta particles and the stripped companion material. In each of our simulations, the supernova phase evolves for approximately two hours of physical time.
The density evolution during this phase for 5m is shown in Fig.~\ref{figure:snr}. 

This distribution (after excising the companion star) is then used as the initial state for the ISM interaction phase of the calculation. 
It is expected that this material will travel unimpeded for $\sim$100 yr before the swept up ISM mass becomes non-negligible, $R_{\textrm{SNR}}\sim$0.13 pc. 
This is roughly the Str\"{o}mgren radius for an 8 \msol\ star, the maximum mass of a white dwarf progenitor star.
During this time, the material will expand considerably in radius and drop significantly in density. 

We begin the ISM phase by expanding the ejecta and stripped material to an initial radius of 0.13 pc and surrounding it with an ambient medium of a uniform density $\rho_{ISM}=1.67\times10^{-24}$ gm cm$^{-3}$, corresponding to a nominal ISM number density of one hydrogen atom cm$^{-3}$.
The ISM particles are distributed using the RPRPS scheme, as was used for the companion and white dwarf. Although the density of each ISM particle is typically much less than an ejecta particle, we enforce particle mass parity between the ejecta and ISM materials. This ensures that any hydrodynamic instabilities that develop are not seeded by mismatched particle masses. The ISM begins at a radius of 0.13 pc and extends to 3 pc, covered by a total of $\approx$ 1.8 million particles. Such a large range ensures that the ejecta and shock are confined within the ISM during the entire evolution of the interaction.

As we are interested in the interaction between the ISM and the ejecta, self gravity of the gas is ignored and both the ISM gas and the supernova ejecta material are modeled with an ideal gas equation of state with $\gamma$=5/3, appropriate for monoatomic gas. 
Each of these simulations has a total simulation time of $\sim$1000 yr. 
In order to compare our results with a companion to those of an isolated supernova explosion, we run an additional ISM phase model with spherically symmetric ejecta material -- in other words, a SNe Ia model consisting of a solitary white dwarf and no companion, hereafter labeled as NC.

\section{Results}

We now turn our attention to the results of each phase of our simulations, starting with the explosion phase and the formation of the hole in the ejecta. 
We then study the formation of the supernova remnant as the ejecta interacts with the interstellar medium. 
Finally, we examine the evidence for the long term survivability of the hole. 

\subsection{Impact of SNR on the Companion Star}
\begin{figure*}
\centering
\setlength{\tabcolsep}{0mm}
\includegraphics[width=0.95\textwidth]{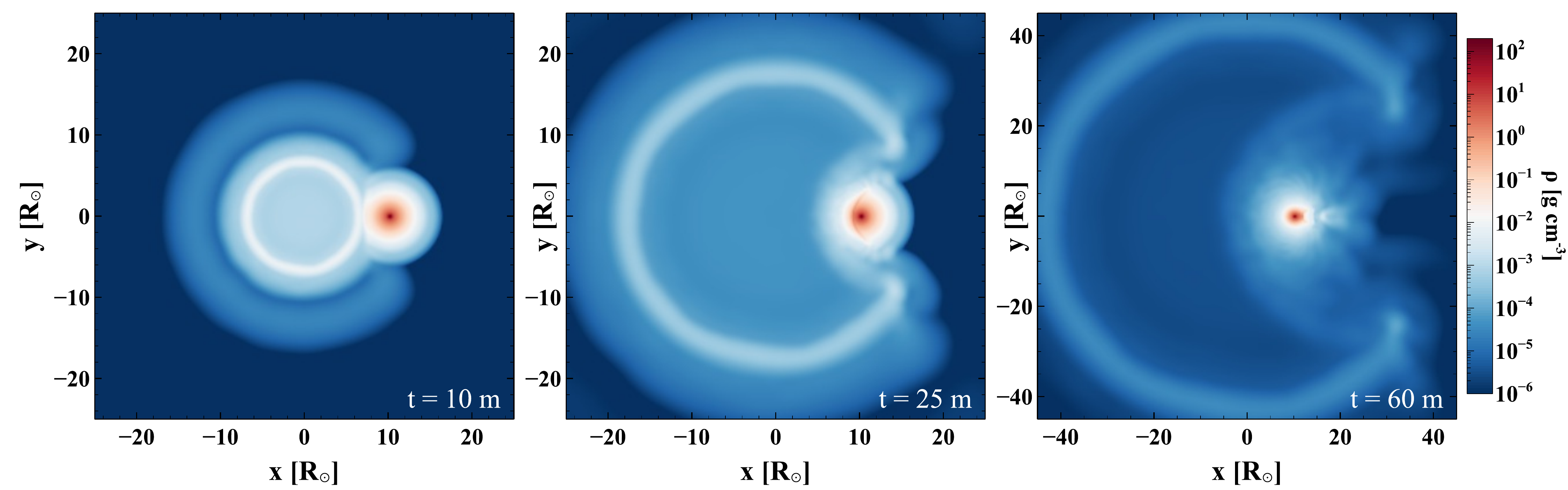} \\
\caption{ Evolution of 5m during the supernova phase. Each panel shows a density slice through the center of the domain with time increasing left to right, corresponding to 10, 25, and 60 minutes after the explosion of the white dwarf. The $x$ and $y$ axis units are given as solar radii. The formation of a hole in the ejecta is clearly seen in the final panel. }
\label{figure:snr}
\end{figure*}

Figure~\ref{figure:snr} shows the density evolution of the supernova and companion star for the 5m run as the supernova remnant develops and overtakes the companion. 
From left to right, these slices correspond to 6, 25, and 60 minutes after beginning of the supernova, respectively. 
The middle and right panels clearly show the formation of the hole by the companion star. At $t$=60 minutes, the hole is roughly cone-shaped with an apex angle of $\sim$40$^{\circ}$ measured from the downstream axis. This is consistent with the results of \cite{Marietta2000} and \cite{GarciaSenz2012} and corresponds to a solid angle of 1.47 sr. 

However, the companion is also altered by this interaction. Although the central density does not change much as the supernova shock runs over the companion, some of its outer layers are stripped off.
Some of this material can be seen on the back side of the companion in the rightmost panel of Figure~\ref{figure:snr}. 

\begin{figure}
\centering
\setlength{\tabcolsep}{0mm}
\includegraphics[width=0.40\textwidth]{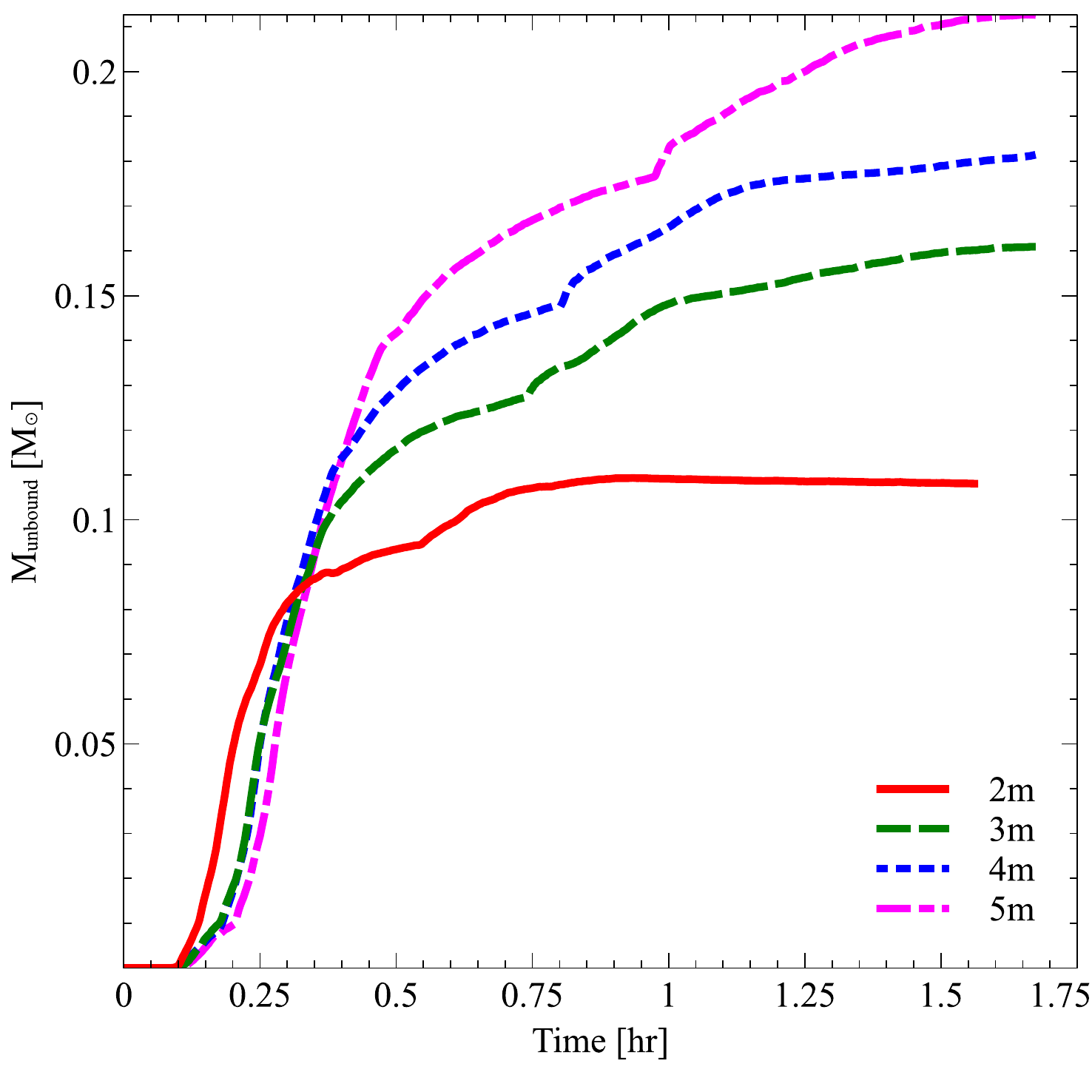} \\
\caption{Plot of the total mass unbound from the companion star for each of our models. The $x$-axis gives the simulation time in units of hours while the $y$-axis is the total stripped mass. }
\label{fig:stripped}
\end{figure}

To estimate the total mass stripped from the companion star, we simply compute $\alpha = T/V$ for every particle in companion, where $T$ is the kinetic energy and $V$ is the potential energy. 
The total stripped mass is then simply the summation of all particles with $\alpha>1$. 
As mentioned above, this stripped material is saved and used a part of the initial conditions during the ISM phase. 
The total mass stripped from the companion for each of our models is given in Table~\ref{table:ic} under the column labeled M$_s$.
Figure~\ref{fig:stripped} shows the total mass stripped from each companion as a function of simulation time. 

We find that roughly 5\% of the companion mass is stripped from the companion star. 
This is slightly lower than that found by \cite{GarciaSenz2012} ($\approx0.1$\msol\ stripped from their 1.0\msol\ companions). 
This is also slightly lower than that of \cite{Marietta2000} wherein they found that between 0.17 and 0.25 \msol\ was stripped from their 1.0\msol\ and 2.0\msol\ red giant companions. However, these differences are subtle considering the range in initial conditions and methods used.

Estimates of the amount of mass stripped from a companion star have been developed by several authors \cite[\eg][]{McCluskey1971,Sutantyo1974A,Sutantyo1974B}, culminating in the theoretical estimates of \cite{Wheeler1975}. 
The total amount of material lost by the companion star is a combination of material directly stripped by the ejecta and material ablated off the star due to the rarefaction wave generated as the ejecta shock runs into the companion. 
Using the procedure found in \cite{Wheeler1975} and using density profiles of our companion stars, we can predict that roughly half the unbound mass is directly stripped and half is ablated from the companion star.
We also find that the total unbound companion gas seen in our simulations agrees very well with the theoretical estimates. 
The estimated theoretical unbounded mass for each of our models is given in Table~\ref{table:ic} under the heading M$_{s,t}$.

\subsection{ISM Evolution}

The large scale structure of the supernova remnant is largely dependent on the interaction between the supernova ejecta and the surrounding interstellar medium (ISM). The structure of the ISM has been studied in \cite{Badenes2006, Badenes2008} which found that a uniform medium with a $\rho_{\rm ISM}$ of 1.67$\times$10$^{-24}$ gm cm$^{-3}$ is a reasonable approximation. This motivates our choice for the ISM density above, \cite[\eg][]{Raskin2013}. The initial radius of the ISM was chosen such that very little of the ISM has been swept up, but future expansion of the ejecta will have an important impact on the ISM. For a spherically symmetric blast wave, the fraction of the ISM swept up is approximated by $f=4\pi\rho_{\rm ISM} R^3/3 M_{\rm SN}$, where M$_{\rm SN}$ is the mass of the ejecta. For our parameters, $f$=0.05\% in the initial condition of the ISM phase. Also, as mentioned above, this radius corresponds to the Str\"{o}mgren radius of the white dwarf progenitor. 
However, it should be noted that as we have simply expanded the ejecta material to this radius in an offline fashion, the mass of swept ISM material included in the simulation is actually zero. 

\begin{figure*}
\centering
\setlength{\tabcolsep}{0mm}
\includegraphics[width=0.95\textwidth]{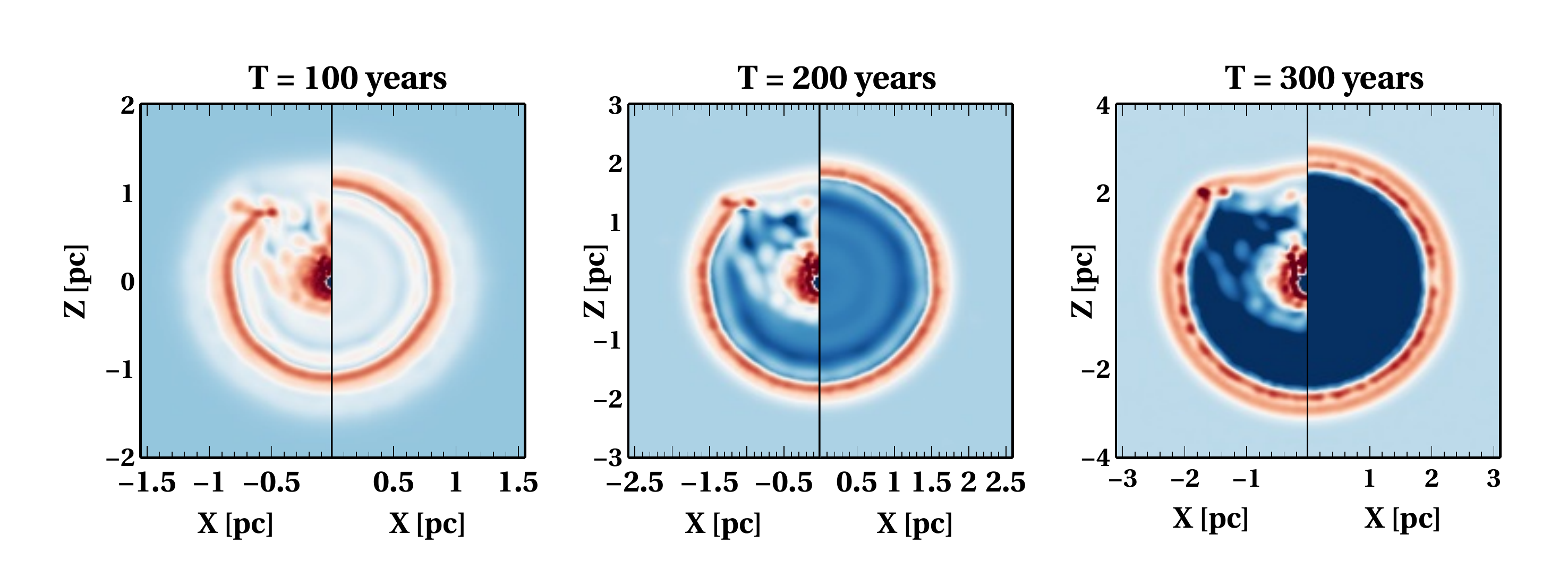} \\
\caption{ Evolution of the ISM phase at {\it Left Panel:} 100 yr, {\it Middle Panel:} 200 yr and {\it Right Panel:} 300 yr. For each panel 5m is shown on the left while NC is shown on right. }
\label{figure:ism}
\end{figure*}

Figure~\ref{figure:ism} shows the density evolution as a slice through the center of the ejecta for 5m at three distinct times as the ejecta interacts with the ISM. The left side panels show 5m whereas the right side panels show a model without a companion star. The hole in the ejecta is readily seen after 100 and 200 years of evolution. This suggests that although some of the material stripped from the companion is found behind the companion, it is not sufficient to fully fill in the hole before it reaches the ISM. By 300 years, some of the swept up ISM material can be found covering the hole in the ejecta. However, this material is also unable to fill in the hole and, as we will show below, does not appreciably change the observational signature of the hole. Finally, a substantial amount of ablated material is found at the center of the cavity evacuated by the ejecta. This gas is completely unbound, but is moving roughly ten times slower than the ejecta at the beginning of the ISM phase. While the ejecta material is slowed during its interaction with the ISM, it nevertheless evolves much faster than the stripped material which does not contribute to the global remnant evolution.

We note here an important difference in the ISM evolution between the models presented here and those presented in \cite{GarciaSenz2012}. 
In the previous work, \cite{GarciaSenz2012} found that as soon as the ejecta shock swept up a nearly equal amount of mass in the ISM, large scale Rayleigh-Taylor (RT) instabilities formed. 
As discussed in \cite{Dwarkadas2000}, these instabilities are generated when the shock from the ejecta has swept up a comparable amount of mass from the ISM. This causes the expanding ejecta shock to slow down and generate a reverse shock that moves inward toward the origin of the supernova. This sets up a condition where the pressure and density gradients point in opposing directions and the gas becomes RT unstable.  However, the simulations of \cite{GarciaSenz2012} and SPH simulations in general suffer from numerical instabilities when particles with large mass differences interact. These numerical instabilities then mimic physical mixing events and lead to incorrect results. This situation is avoided in the simulations presented here since an identical particle mass is used for both the ejecta and the ISM. However, there is some indication that RT instabilities are nevertheless forming in our models, specifically in the second and third panels of Figure~\ref{figure:ism}.

\begin{figure}
\centering
\setlength{\tabcolsep}{0mm}
\includegraphics[width=0.40\textwidth]{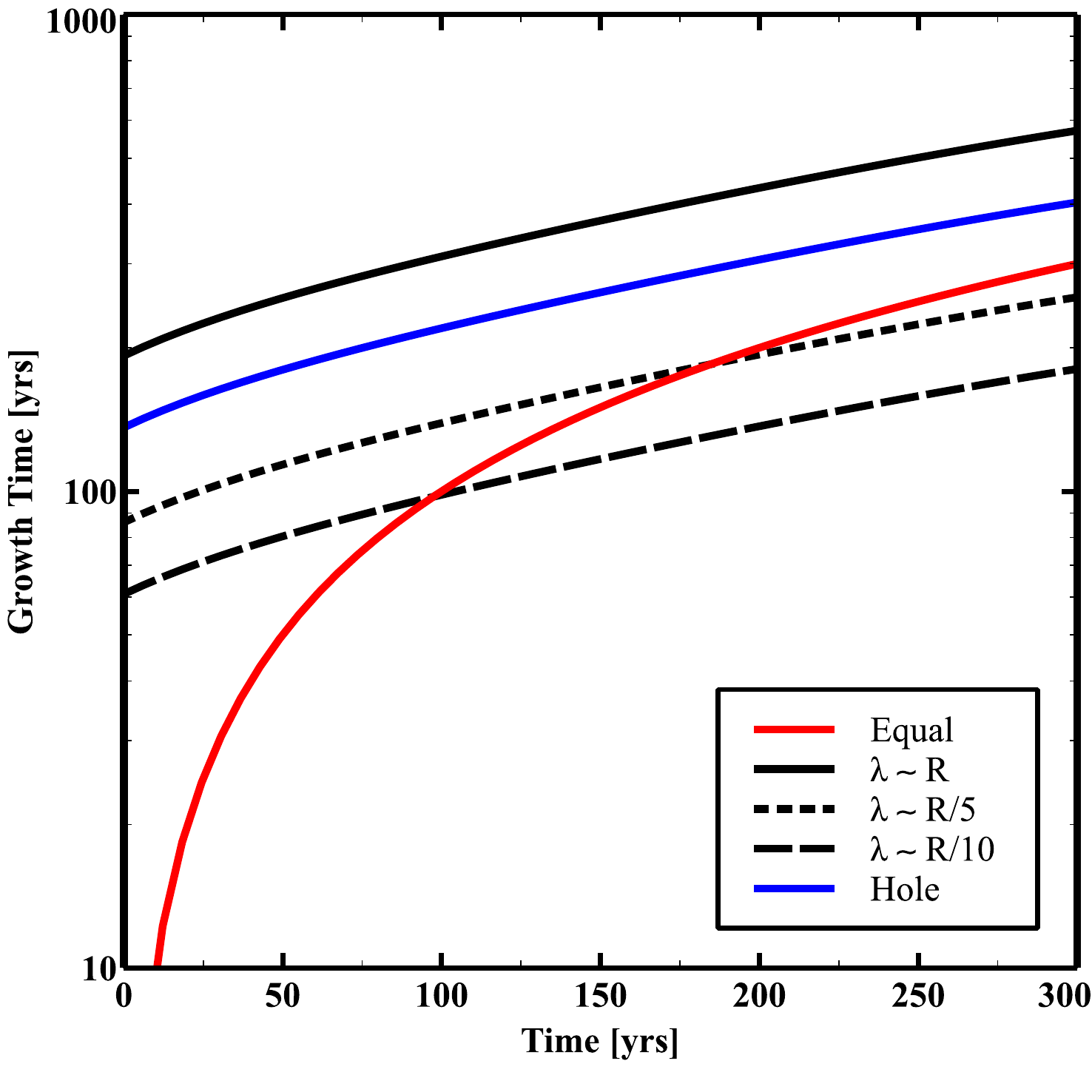} \\
\caption{Estimated Rayleigh-Taylor instability growth rates. The $x$-axis give the time since the beginning of the interaction of the ejecta with the ISM in units of years while the $y$-axis gives the growth time in years. The (red) solid line shows where the growth time is equal to the evolution time. The black lines show the growth time for a range of feature sizes. The (blue) solid line gives the estimated growth time for a feature of equal size to the mass-shadow hole. $\lambda=R/10$ gives a resonable estimate for RT instabilities to grow due to the resolution used. }
\label{figure:grate}
\end{figure}

\begin{figure}
\centering
\setlength{\tabcolsep}{0mm}
\includegraphics[width=0.40\textwidth]{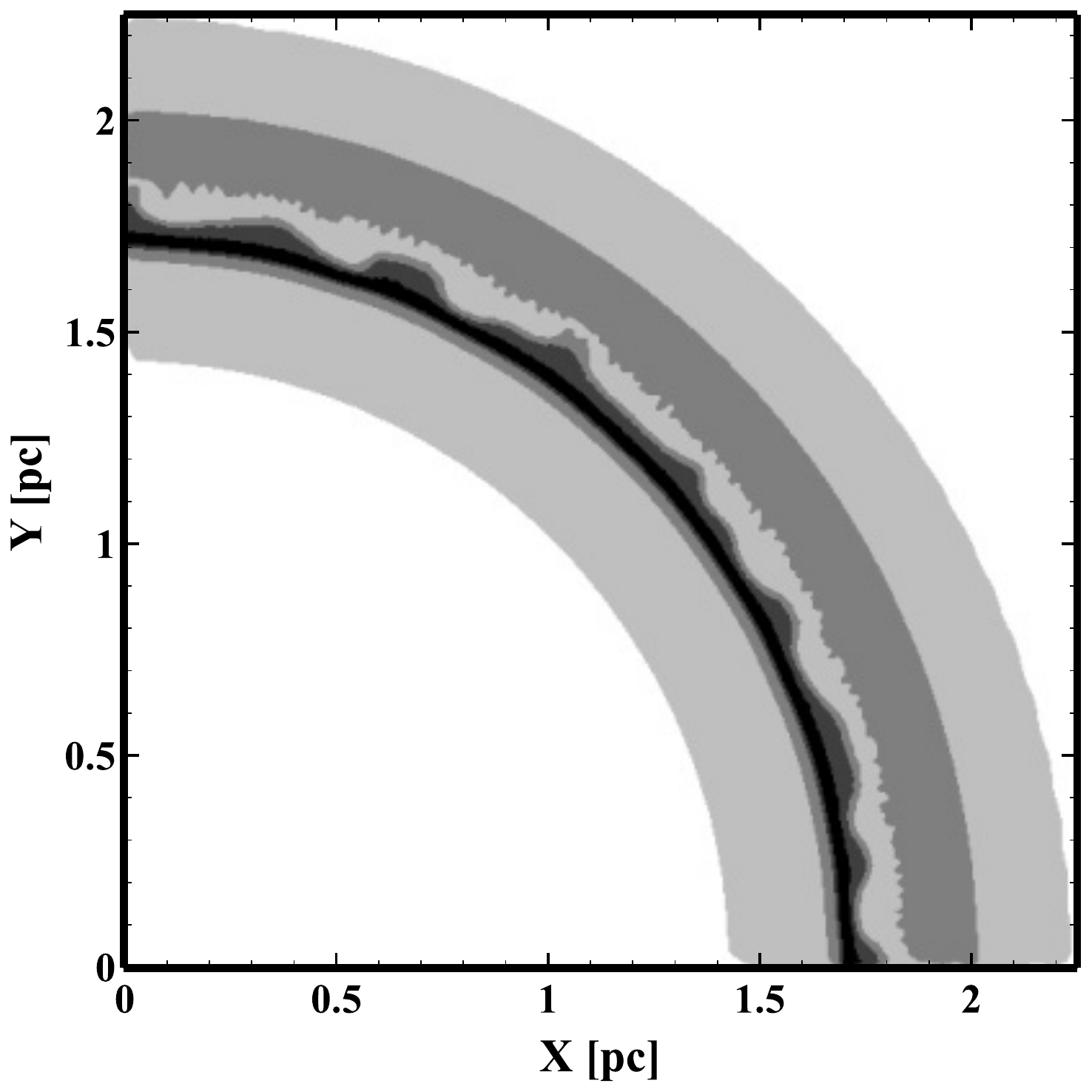} \\
\caption{Density contours in a slice of a high resolution octet at $t$=200 years showing the development and growth of Rayleigh-Taylor instabilities. These instabilities are found at many length scales.}
\label{figure:rt}
\end{figure}

The growth rate of the RT instability can be estimated using the physical properties of the ejecta and the ISM. The classical growth rate of RT instabilities in the linear regime is given by 
\be
\Gamma = \sqrt{ A_t a_{s} k},
\ee 
where $A_t = ({\rho_{H}-\rho_{L}})/({\rho_{H}+\rho_{L}})$ is the Atwood number which is a dimensionless density ratio determined by the shock density and the ambient density, $a_{s}$ is the acceleration of the discontinuity, and $k=2\pi/\lambda$ is the spatial wavenumber. The Atwood number and radius of the shock are measured at several times from the model without the companion, noted above as NC. $a_s$ is then measured by assuming the shock is moving ballistically and fit by $R= R_0 + v_0t + 1/2 a_s t^2$. Figure~\ref{figure:grate} shows the estimated growth rate for a range of instability sizes. In particular, the blue line shows the growth time for an RT instability of comparable size to the mass-shadow hole. It is clear that RT instabilities on this scale do not grow fast enough to fill in the hole before it can be observed. The features found in the second and third panels of Figure~\ref{figure:ism} are seeded by the numerical shot noise of the ISM particle distribution.

To validate these growth estimates, we ran a set of high resolution models to study the development of the RT instability. For numerical convenience, these models are limited to a single octet. The supernova ejecta density and velocity profiles are spherical averages of the profiles from the NC supernova phase, and we generate particle positions using RPRPS. \cite{Raskin2016} have demonstrated that this method results in less particle shot noise than other distribution approaches, but does not completely eliminate noise. It is this noise that seeds the growth of the RT instability. 

Figure~\ref{figure:rt} shows the development of the RT instability after 200 years of evolution for a simulation with a total of 2.3 million particles. RT instabilities have begun to grow over the surface of the ejecta as it moves into the ISM. In fact, RT instabilities are forming at many disparate length scales with growth times consistent with the estimated growth rates given shown in Figure~\ref{figure:grate}, with small to medium length instabilities beginning to grow, while larger instabilities that might close the hole have yet to develop.

\subsection{X-Ray Emission}

The thermal x-ray emission is generated by bremsstrahlung (free-free emission), recombination (free-bound emission), and two-photon emission \cite[\eg][]{Kaastra2008, Vink2012}. This x-ray emission is primarily a function of the electron density, temperature, and the ion density and thus is roughly dependent on the square of the local density, $\propto\rho^2$. Since x-rays are limited to energy ranges between 100 eV and 100 keV, only elements with large binding energies can produce these photons. In the analysis that follows, we consider only the ejecta particles as they are initially composed of carbon and oxygen. To estimate the x-ray emission along a particular line of sight, we use a series of ray trace calculations through the center of the ejecta. 

\begin{figure}
\centering
\setlength{\tabcolsep}{0mm}
\includegraphics[trim= 60.0mm 25.0mm 25.0mm 25.0mm, clip,width=0.45\textwidth]{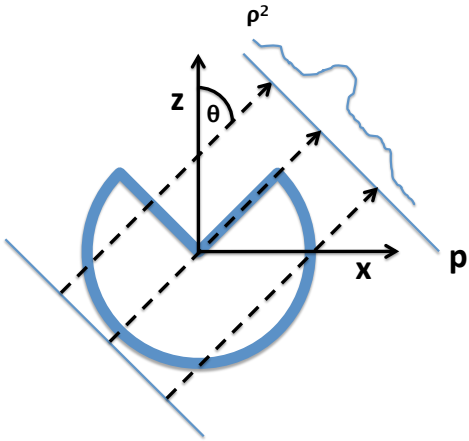}
\caption{ Schematic representation of the ray tracing scheme. The x-ray emitting ejecta with a hole is represented as the blue nearly-spherical shape. The projected coordinate p is orthogonal to the line of sight, defined by the line of sight angle $\theta$. The (black) dotted lines represent the rays used in the procedure and along which the density is summed. }
\label{figure:rtscheme}
\end{figure}

Following the procedure described in \cite{GarciaSenz2012}, we determine the x-ray emission along a particular line of sight by means of a simple ray-tracing routine. 
We compute $\rho_j$ at 500 points along each ray and a total of 500 rays are computed. 
The total value for $\rho^2$ is computed by summing up the square of the density along the ray. 
Additional lines of sight are found by rotating the particle data and recasting the ray. 
Different lines of sight are then simply described by the angle $\theta$ by which the data was rotated. 
In computing the x-ray emission, only particles with temperatures greater an 10$^{6}$ K are considered.
These ray traces are computed at several times during the ISM phase to study the evolution of the hole. 
Therefore, the physical volume that is probed by the ray trace changes with time. 
To facilitate a systematic comparison between models and time, we plot these ray traces in terms of the projected coordinate. 
Schematically, this is shown in Figure~\ref{figure:rtscheme}. This figure is similar to that of \cite{GarciaSenz2012}. 

\begin{figure}
\centering
\setlength{\tabcolsep}{0mm}
\includegraphics[width=0.40\textwidth]{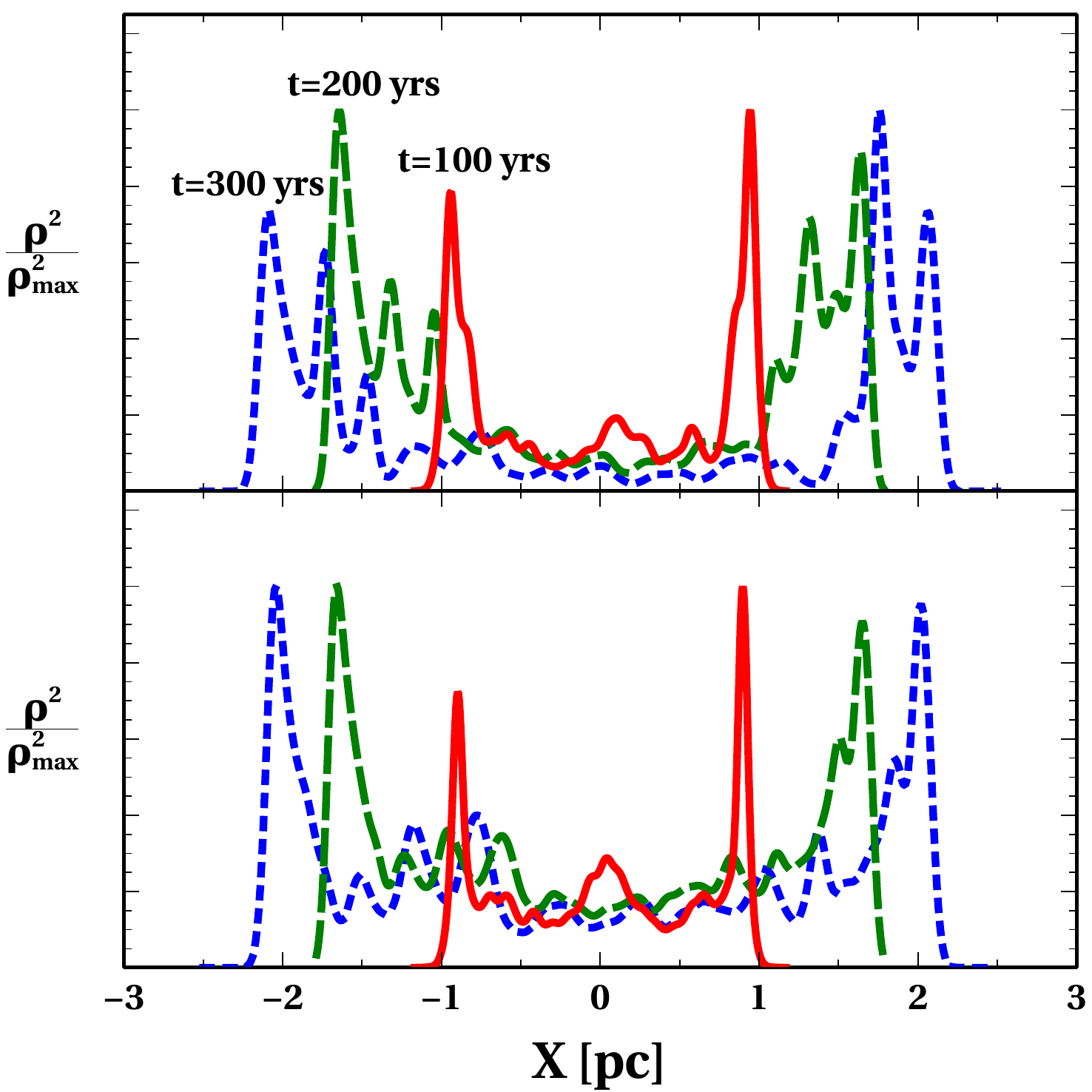} \\
\caption{ {\it Top Panel: } $\rho^2$ profiles of 5m as a function of time. {\it Bottom Panel:} $\rho^2$ profiles of NC. The solid (red) lines shows $t$=100 yrs after beginning of the ISM phase, the dashed (green) line shows $t$=200 yrs, and the dotted (blue) line shows 300 yrs. The $y$-axis is $\rho^2$ normalized by the maximum value of $\rho^2$. The $x$-axis is the projected axis in physical units orthogonal to the light of sight with $\theta$=0.}
\label{figure:vtime}
\end{figure}

The top panel of Figure~\ref{figure:vtime} shows the $\rho^2$ profiles for 5m at several times during the ISM phase while the bottom panel shows the evolution of NC, the same model as 5m but without the companion star and represents an isolated spherically-symmetric supernova. Each line is normalized to unity. These profiles are computed along a line of sight that looks directly down the hole formed by the companion. This corresponds to a line of sight angle of $\theta$=0. 

The edge of the ejecta is easily seen as the set of prominent peaks in both panels of Figure~\ref{figure:vtime}. The presence of the hole created by the companion is also evident as the set of smaller secondary peaks found near the primary peaks. These secondary peaks are found throughout the ISM phase, becoming even more prominent at later times. By the final time, these secondary peaks are nearly as large as the peaks generated by the ejecta shock. The slight peaks found in the middle of the $t=100$ years is due to the still relatively dense and hot ejecta material. Finally, even the symmetric case begins to develop secondary peaks at late times. There are two possible explanations for these peaks. First, since we do not account for any cooling with the ejecta shock, this maybe due to the ejecta shock smearing out over a larger region than it would with cooling. Second, it maybe due to RT instabilities mentioned above. 

\begin{figure}
\centering
\setlength{\tabcolsep}{0mm}
\includegraphics[trim= 00.0mm 00.0mm 00.0mm 00.0mm, clip, width=0.40\textwidth]{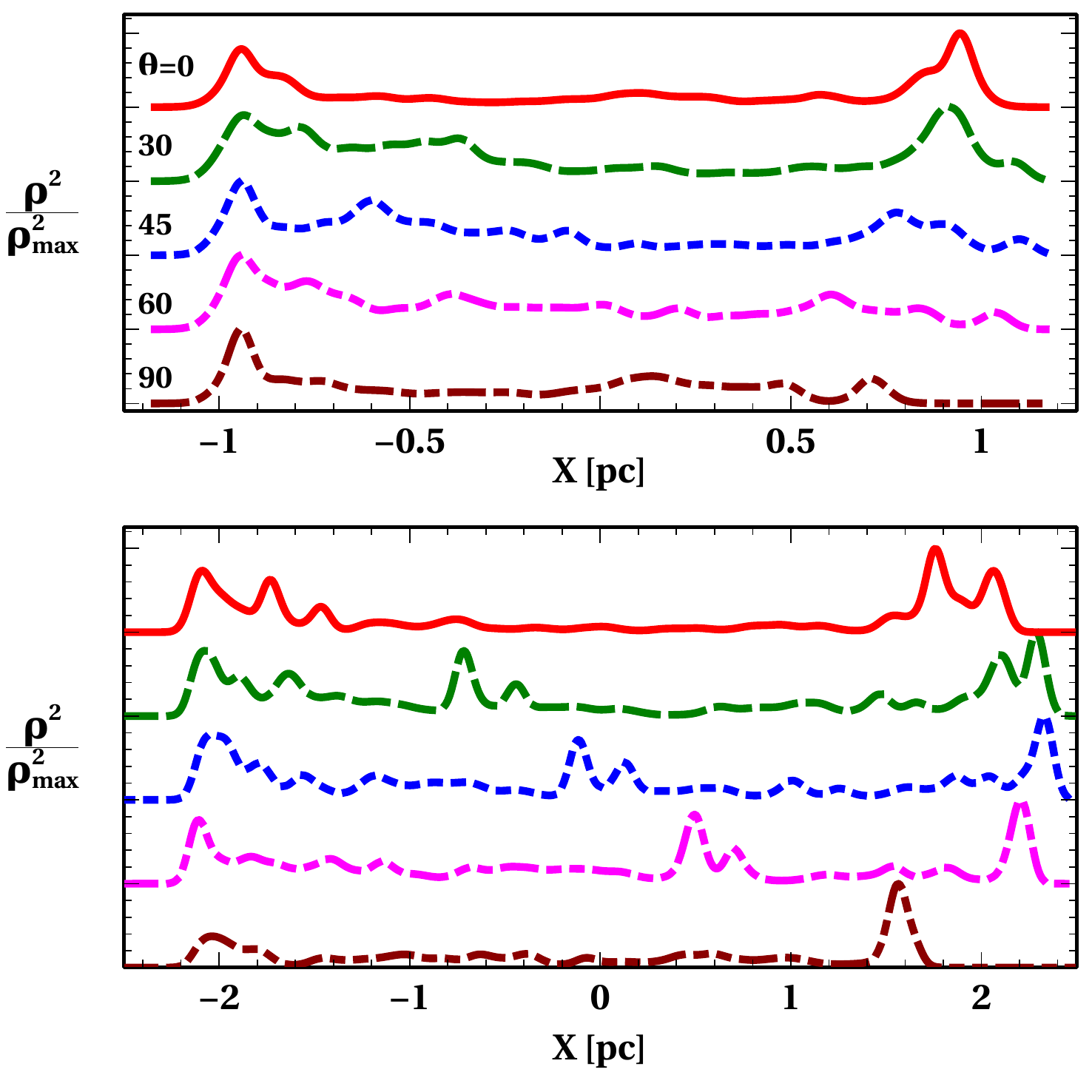} \\
\caption{Normalized $\rho^2$ profiles for 5m at {\it Top panel:} 100 years and {\it Bottom panel:} 300 years. The (red) solid line represents $\theta$=0, the (green) dashed line represents $\theta$=30$^{\circ}$, the (blue) dotted line shows $\theta$=45$^{\circ}$, the (magenta) dash-dotted line shows $\theta$=60$^{\circ}$, and the (dark red) shows $\theta$=90$^{\circ}$. An offset is used to make each line distinct.}
\label{figure:vangle}
\end{figure}

Figure~\ref{figure:vangle} shows a comparison of viewing angles at $t=100$ years (top panel) and $t=300$ years (bottom panel) for 5m. Here we compute additional lines of sight that correspond to a range of $\theta$ between 0 and 90$^{\circ}$. As mentioned above to simulate these additional lines of sight, we have rotated the original SPH data around the $y$-axis by an angle $\theta$. As shown in Figure~\ref{figure:snr}, the ejecta material near the hole is slightly higher in density than compared to rest of the ejecta. This density enhancement plays an important role at early times during the ISM phase.

At $t$=100 years, the edge of hole created by the companion is clearly present in the $\theta$=0 cases as the two secondary peaks near the primary peaks created by the ejecta. As $\theta$ increases, however, one side of the hole becomes coincident with the edge of the ejecta while the other creates an increase in $\rho^2$ inward of the other side of the ejecta. In the $\theta$=30 case, this is apparent as the extended emission around $x\sim$-0.75. At larger angles the effect of the hole is more evident as a lack of emission near the $x$=1.0. Although the hole is slightly denser on average than the rest of hole, its path length is much shorter. This leads to a decrease is the estimated emissivity of the x-rays and a decrease in $\rho^2$. At $\theta$=90, this effect is easily seen as the clear asymmetric peaks.

At $t$=300 years, this effect is more drastic. At $\theta$=0, the ejecta and hole form a clear series of peaks. As $\theta$ increase, there is a clear signal from one edge of the hole as it moves from one extreme towards the center of the projection. Between $\theta$=30 and $\theta$=60, the emissivity from the hole edge is nearly as strong as the signal from the ejecta itself. At $\theta$=90, however, the density of the hole has fallen less than the rest of the ejecta. This produces an enhancement in the emissivity when compared to the ejecta. This causes a clear asymmetry that is opposite to that found at $t$=100 years. 

\begin{figure*}
\centering
\setlength{\tabcolsep}{0mm}
\includegraphics[trim= 00.0mm 10.0mm 00.0mm 00.0mm, clip, width=0.85\textwidth]{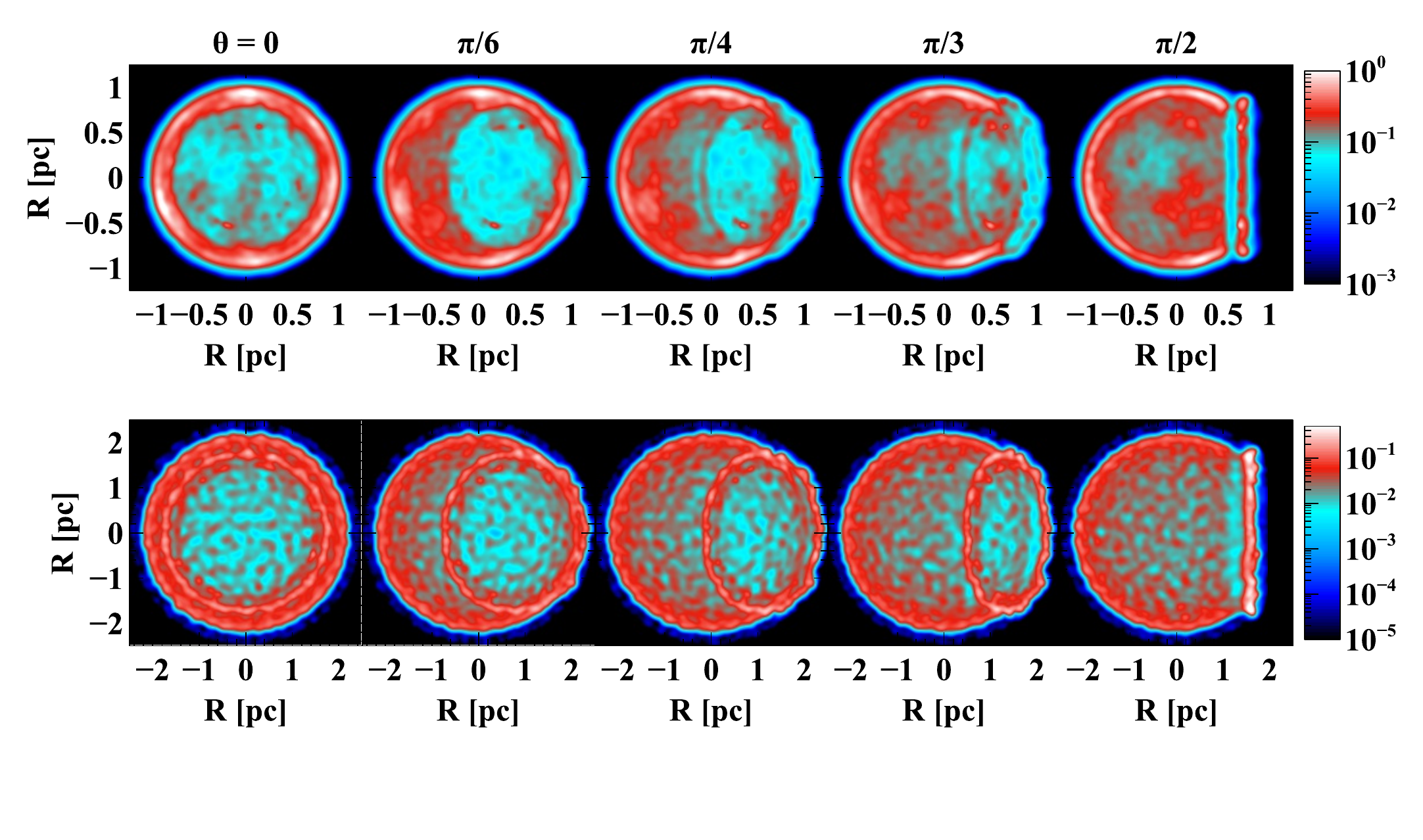} \\
\caption{2D projected images of the derived x-ray emission for 5m {\it Top Row:} $t$=100 years and {\it Bottom Row:} $t$=300 years. Each row shows a different viewing angle given as $\theta$ in the title. Each figure is normalized by the maximum value in the $\theta$=90 case.}
\label{figure:2dxray}
\end{figure*}

\begin{figure}
\centering
\setlength{\tabcolsep}{0mm}
\includegraphics[width=0.40\textwidth]{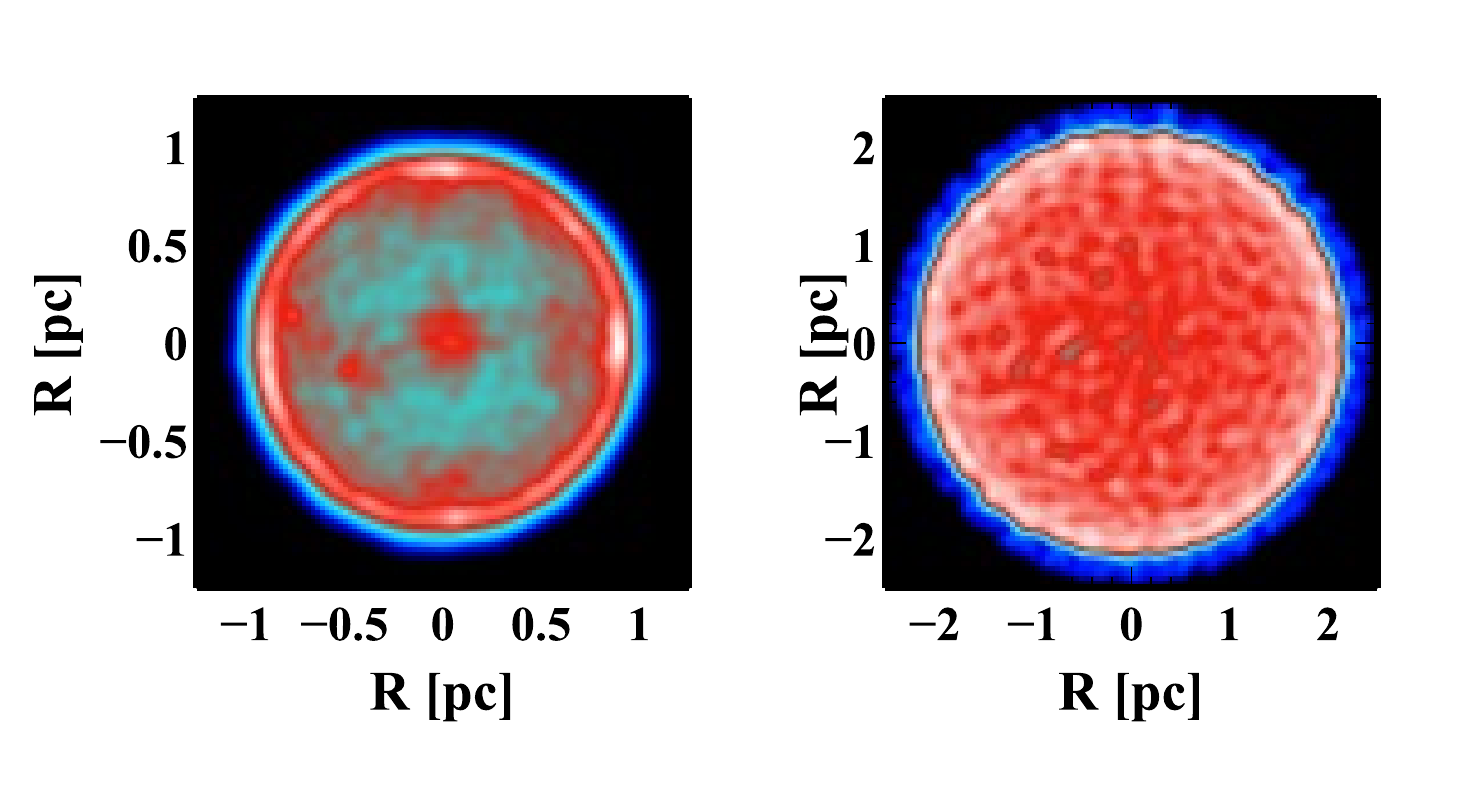} \\
\caption{2D projected images of the derived x-ray emission for {\it Left Panel:} $t$=100 years and {\it Right Panel:} $t$=300 years for NC.}
\label{figure:2dxrays}
\end{figure}

Finally, Figure~\ref{figure:2dxray} shows a two dimensional approximation of the x-ray emission as a function of time and viewing angle. The top row shows $t$=100 years while the bottom row shows $t$=300 years. The presence of the hole created by the companion is clearly seen at both times but is most prominent for $t$=300 years. At $\theta$=90 the hole creates a clear asymmetry in the emission that is present at both times. For comparison the simulated x-ray emission for NC is shown in Figure~\ref{figure:2dxrays} at $t$=100 and $t$=300 years. 

\begin{figure}
\centering
\setlength{\tabcolsep}{0mm}
\includegraphics[width=0.40\textwidth]{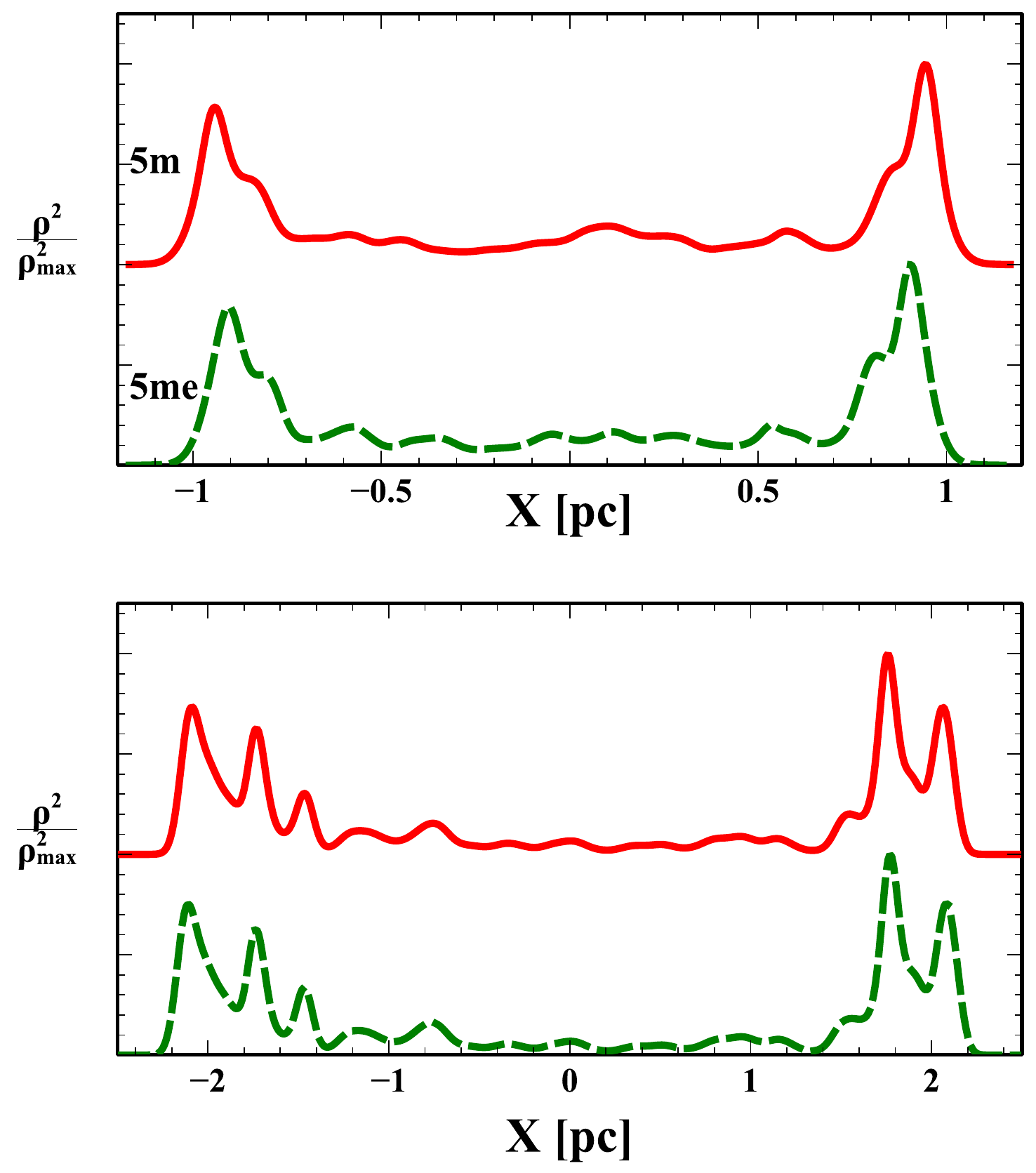} \\
\caption{ Normalized $\rho^2$ profiles for 5m and 5me at {\it Top panel:} 100 years and {\it Bottom panel:} 300 years. The (red) solid line shows 5m and the (green) dashed line shows 5me, a model where the stripped companion material is ignored.}
\label{figure:5mcomp}
\end{figure}

Figure~\ref{figure:5mcomp} shows a comparison between 5m and a model where we have ignored the effect of the stripped material from the companion in the hydrodynamics during the ISM phase, denoted as 5me. The top panel shows $t$=100 years while the bottom shows $t$=300 years. Although only the ejecta material is considered in computing $\rho^2$, it is apparent that the stripped material plays no role in either the hydrodynamics or observability of the hole. In fact, the profiles shown in Figure~\ref{figure:5mcomp} are essentially identical at both 100 and 300 years. 

\begin{figure}
\centering
\setlength{\tabcolsep}{0mm}
\includegraphics[width=0.40\textwidth]{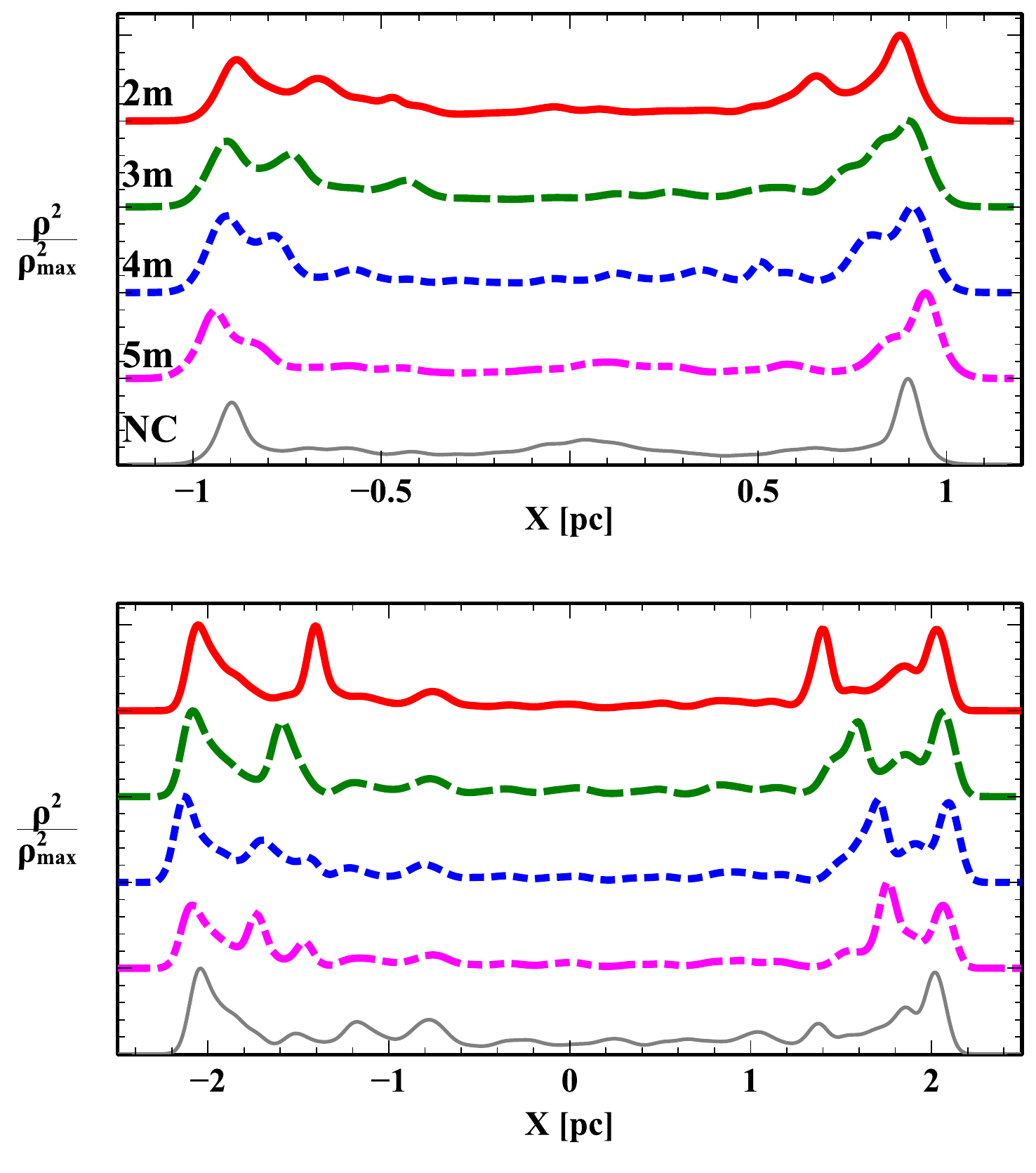} \\
\caption{ Comparison of normalized $\rho^2$, $\theta$=0 profiles for a range of companion masses at {\it Top panel:} 100 years and {\it Bottom panel:} 300 years. The (red) solid line shows profiles for 2m, the (green) dashed line shows 3m, the (blue) dotted line shows 4m, the (magenta) dash-dotted line shows 5m, and the (gray) thin solid line shows NC. }
\label{figure:mcomp}
\end{figure}

Next we turn our attention to the relationship between the initial companion mass, and therefore size, and the final size of the hole which is shown in Figure~\ref{figure:mcomp}. As mentioned above, although the physical size of each companion varies greatly, the solid angle that each covers in relation to the white dwarf is roughly the same. Consequently, although the 2\msol\ companion is less than half the size of the 5\msol\ companion the size of the hole generated is nearly the same. Only a slight dependence on the mass of companion where the hole generated in the ejecta is more pronounced with smaller mass companions. In fact, at early times the difference in $\rho^2$ due to companion mass is nearly zero. Only at late times is it possible to differentiate between the companion masses. In general the distance between the two peaks in $\rho^2$ is inversely proportional to the companion mass with the largest separation found in 2m. Finally, we also compare $\theta$=0 profiles for NC, the model without a companion. The presence of the hole is easy to see as the extended emission near x=0.6-0.9 at $t=$100 yrs that is absent in NC. At $t$=300 yr, the hole is much more evident as the dominant set of peaks in the emission that is not found in NC.

\section{Resolution Study}

\begin{figure}
\centering
\setlength{\tabcolsep}{0mm}
\includegraphics[width=0.40\textwidth]{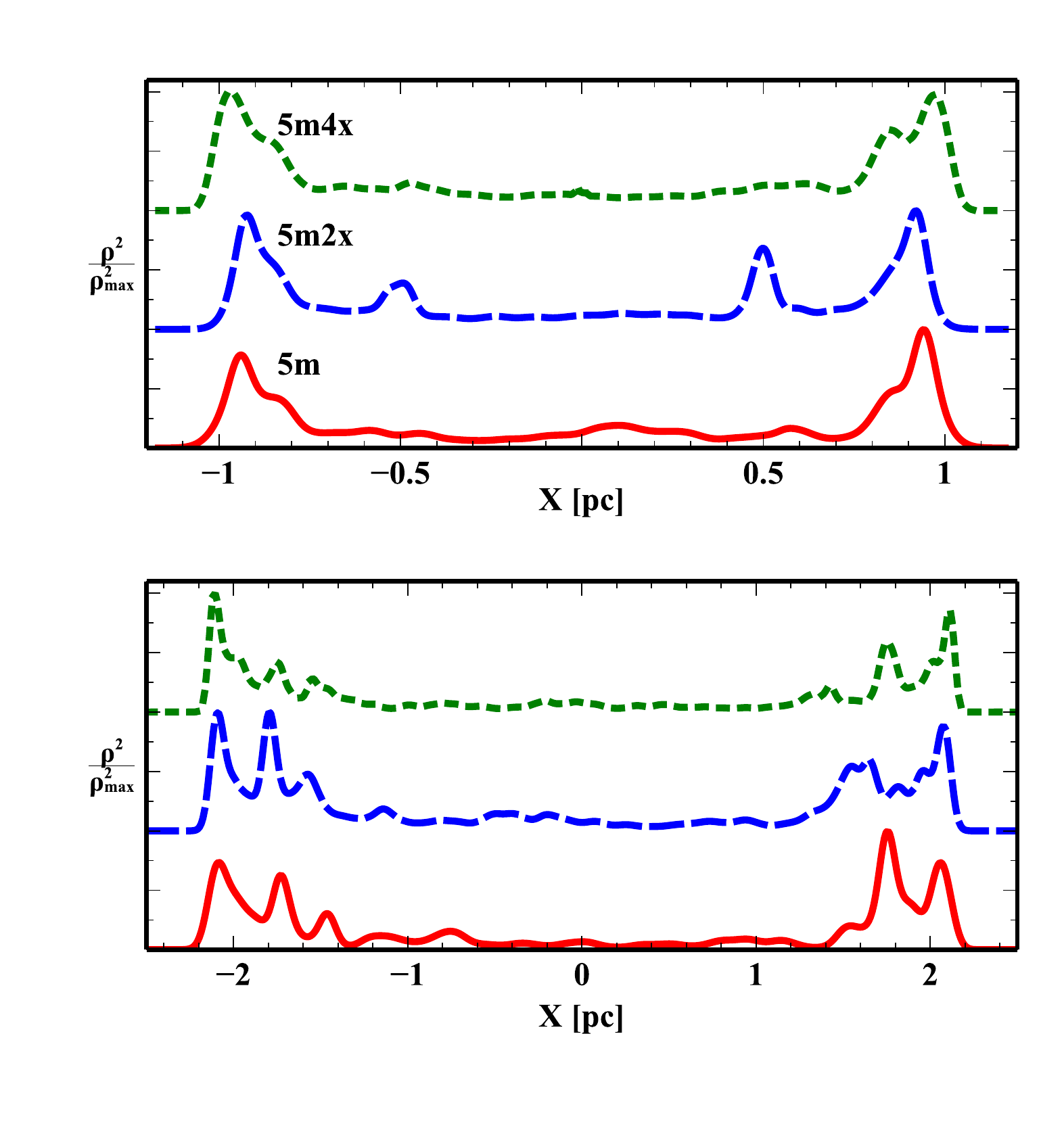} \\
\caption{ Comparison of normalized $\rho^2$ profiles for a range in particle mass at {\it Top panel:} 100 years and {\it Bottom panel:} 300 years.The solid (red) line shows 5m, the (blue) dashed lines shows 5m2s, and the dotted (green) line shows 5m4x.}
\label{figure:resolution}
\end{figure}

Finally, we explore the effect of resolution on our models. We have repeated 5m two additional times with twice (denoted 5m2x) and four times (denoted 5m4x) the nominal particle resolution with a total particle count of 900k and 1.8 million particles respectively, during the supernova phase.  Figure~\ref{figure:resolution} shows the estimated x-ray emission at 100 and 300 years after the beginning of the ISM phase for 5m, 5m2x, and 5m4x.

The appearance of secondary peaks in the 2x resolution study are the result of an overdense ring of material around the opening of the hole. This ring forms as the reverse shock from the shell encounters the hole edge and travels transversely to the outward flow of material. There is some subtle indication of this happening in the 4x case as well, but the density is not as strongly peaked at the hole edge. We suspect this simulation has simply not converged on a single behavior regarding the over density around the hole edge. However, this has little bearing on the conclusions presented here as the bulk evolution of the remnant and the preservation of the hole are features common to all realizations.

However, by 300 years, all three models look very similar. Although the strength of each peak varies slightly due to the resolution, the bulk properties remain the same. That is, the presence of the hole is still seen as the set of secondary peaks at 300 years. In general, the effect of resolution is minor.

\section{Summary \& Conclusion}

In the single degenerate SNe Ia scenario, the collision between the supernova ejecta and the companion is unavoidable. This has prompted theoretical studies of this interaction as a method of identifying companion properties. While \cite{Kasen2010} computed a series of light curves that suggested this interaction should be visible in $\sim$10$\%$ of all single degenerate SNe Ia, \cite{Kutsuna2015} showed that this emission was highly dependent on the initial separation between the companion and the white dwarf. They established a strict cutoff at $\sim$2$\times$10$^{13}$cm, below which such emission would be invisible. \cite{Meng2016} showed that many existing binary systems have separations below this cutoff. 

On the other hand, a hole in the supernova ejecta may be observable at late times depending on the evolution of the ejecta as it interacts with the interstellar medium. \cite{GarciaSenz2012} found in their axisymmetric simulations that the hole in the ejecta persisted for many centuries after the interaction with the interstellar medium. However, they also found that over such long timescales, hydrodynamic instabilities at the edges of the hole may eventually close the hole. 

We have presented here a set of high resolution 3D SPH simulations that aim to illustrate the effect of a companion star on the evolution of a supernova remnant and the observability of the hole produced from the interaction with the companion. Our simulations have two distinct advantages over previous studies in this area. First, all of models are fully 3D to capture the full extent of the degrees of freedom during the interaction. Second, all of the particles have identical mass, avoiding possible numerical artifacts. 

Each of the simulated companion stars were early stage red giant branch stars that varied in mass from 2\msol\ to 5\msol\ while the supernova was modeled as a 1.0\msol\ white dwarf with an initial explosion energy of $\approx$10$^{51}$ ergs. For each of the simulations, we find that between  0.11-0.22 \msol\ of material was stripped, depending on the companion mass, matching the estimates from \cite{Wheeler1975}. 

We have presented estimates of the x-ray emission profiles over a wide range of times, companion masses, and angles. The signal from the hole is found as a secondary set of peaks in the x-ray emission. In the head-on $\theta=0$ case, these peaks are found near the primary peaks from the ejecta shell. At larger angles these peaks are found separate and distinct from the ejecta shell. At $\theta=90$ the emission takes on an unmistakably asymmetric appearance.  We find that the hole is present at all viewing angles both early ($t=100$ years) and late ($t=300$ years) after the interaction with the ISM. Finally, we find the hole should remain over the long term evolution of the supernova remnant. 

In a future work, in addition to increasing the resolution, we plan on improving the models presented here. One important feature to include is to use a more realistic ISM. The ISM was modeled here as a uniform medium with constant density. In reality, the ISM is much more complicated and composed of different phases \cite[\eg][]{Cox2005}. The inclusion of nuclear burn network in order to track the formation of important radioactive elements would also improve the estimated x-ray emission. Furthermore, as we have neglecting any cooling physics in the ISM interaction phase, our density profiles are too smooth. Allowing for radiative cooling after the Sedov phase might produce an even stronger signal of the remnant hole. Finally, the inclusion of more realistic treatment of the emission is warranted to validate the results of the estimated x-ray emission.

The results presented here provide an essential tool in the study of supernovae remnants. In particular, we have shown that the interaction between the supernova ejecta and the non-degenerate companion leaves a long-lived imprint on the final supernova remnant. 

We would also like to thank the anonymous referee for their comments which helped to improve this paper. 
This work was performed under the auspices of the U.S. Department of Energy by Lawrence Livermore National Laboratory under Contract DEAC52-07NA27344.
The authors also acknowledge the Livermore Computing Center at Lawrence Livermore Nation Laboratory for providing HPC resources that contributed to the results reported within this paper. LLNL-JRNL-697001.

\bibliographystyle{apjsingle}
\bibliography{ms.bib}
\end{document}